\documentstyle[11pt,newpasp,twoside,epsf]{article}
\def\edcomment#1{\iffalse\marginpar{\raggedright\sl#1\/}\else\relax\fi}
\reversemarginpar

\begin{document}
\title{
Galaxy Formation and 
Chemical Evolution in Hierarchical Hydrodynamical Simulations}
 
\author{Cora, S. A.}
\affil{ Observatorio Astron\'omico de La Plata, 
Paseo del Bosque s/n, 1900 La Plata, Argentina.}
\author{Mosconi, M. B.}
\affil{
Observatorio Astron\'omico de C\'ordoba, Argentina.}
\author{Tissera, P. B.}
\affil{
Instituto de Astronom\'{\i}a y F\'{\i}sica del Espacio,
Buenos Aires, Argentina.}
\author{Lambas, D. G.}
\affil{
Observatorio Astron\'omico de C\'ordoba, Argentina.}

\begin{abstract}
We report first results of an implementation of a
chemical model in a
cosmological code, based on the Smoothed Particle
Hydrodynamics (SPH) technique.
We show that chemical SPH simulations are a promising
tool to
provide clues for the understanding of the chemical
properties of galaxies
in relation to their formation and evolution in a
cosmological framework.
\end{abstract}

\section{Introduction}

Recent observational data of the deep surveys
(Canada-France Redshift Survey, Lyman-break galaxy surveys, the
Hubble Deep Field) 
have provided information on the astrophysical properties of 
galactic objects at different stages of evolution of the 
Universe, making possible to carry out a suitable confrontation 
of different scenarios of structure formation and
galaxy chemical evolution.
Analytical models of chemical evolution are 
a very powerful tool to study galaxy formation (e.g., 
Chiappini et al. 1997 and references therein).
However, they are 
restricted by several hypothesis (no
inflows or outflows, instantaneous recycling, etc.), and 
cannot include dynamical and kinematical evolution
of the matter according to its nature (dark matter, baryons) in 
consistency with a cosmological model.
Steinmetz \& Muller (1994) have implemented chemical enrichment in a
code based on 
SPH for the first time (see also
Raiteri, Villata \& Navarro 1996).
These works  run prepared-cosmological
initial conditions where the formation and
evolution of one single object is studied.
                                              
Fully-consistent cosmological simulations have the
advantage of providing a coherent well-described
environment for all objects and a complete record
of their formation and evolution.
We report here results of a chemical 
model 
implemented
in a fully cosmological SPH code in hierarchical clustering scenarios.

\section{Chemical Implementation and Results}

We have developed a model                   
to implement metal enrichment in a
cosmological context
based on the AP3MSPH code described by Tissera et al. (1997).

A star formation (SF) algorithm has been included
based on the Smidth law. 
Cold  and dense gas particles that 
satisfy the Jean's instability criterium are eligible to form stars.
Each star cluster formed in a given baryonic particle 
at a SF episode is given by
$ \Delta_{\rm star}=C\,\, {\rho_{\rm gas}^{3/2}}\,\, \Delta t$, 
where $\Delta t$ is the integration time-step,
$\rho_{\rm gas}$ the gas density of the 
particle and $C$ the SF efficiency parameter.
Baryonic particles
carry out the information of the  different stellar populations formed
and the remanent gas mass (hybrid particles). 
When the gas
reservoir of a particle is depleted, it is transformed in a 
star particle.

Particles 
are initially formed by Hydrogen and Helium
in primordial abundances ($\rm H=0.75$, $\rm He=0.25$).
Metals are produced and ejected to the interstellar
medium at the end
of the life of stars.
Most
of the elements are produced by Type II supernovae
(SNIIe),
except for the iron that is mainly produced by 
Type I supernovae
(SNIe). 
Each $\Delta_{\rm star}$
formed can be followed up in time and
the number of stars of a given mass estimated
by assuming an Initial Mass Function (IMF). We adopted a 
Salpeter IMF with
a lower and upper mass cut-off 
of $0.1$ and $120$ ${\rm M_{\odot}}
$, respectively.
We resort to 
Woosley \& Weaver (1995) 
metal ejecta tables 
for SNeII.
The adopted nucleosynthesis prescriptions for Type Ia SNe 
are taken from Thielemann, Nomoto \& Hashimoto (1993).
Type Ib SNe
are assumed to be half the total
number of SNeI and to produce only iron 
($ \approx 0.3 {\rm M_{\odot}}$ per explosion).
We assume that the life 
time of binary stars
that finish their lives as SNI 
($t_{\rm SNI}$) 
is
$\approx 0.5-1$ Gyr. 
Ejected metals are distributed 
within the neighboring sphere of
the particle where a $\Delta_{\rm star}$ is
formed according to the SPH technique.
Hydrogen and
Helium are proportionally decreased
according to the metal mass received 
by the particle
(see for details Mosconi et al. 2000).

Given an IMF, the free parameters of our chemical model are
$C$,
the relative rate of different types of supernovae 
($\Theta_{\rm SN}$=SNRII/SNRI) and  
$t_{\rm SNI}$.
The effects of thermal or kinetic
energy injection into the ISM due to SN explosions
are not included in this work.

We performed SPH simulations consistent
with a Cold Dark Matter (CDM) spectrum with $\Omega =1$,
$\Lambda =0$,
$\Omega_{\rm b}= 0.1$, and
$\sigma_{8}=0.67$.
We used  $N = 262144$
particles (${\rm M_{part}=2.6 \times 10^{8}  M_{\odot}}$) in a
comoving box of length $L=5 h^{-1} $ Mpc ($H_{0}=100 h^{-1}\ {\rm{ km \
s^{-1}\ Mpc ^{-1}}}$, $h=0.5$).
The simulations performed (S1,
S2, S3, S4)
share the same initial conditions but have different SF and SN parameters 
(see Figure 1 for values of these parameters).

We identify galaxy-like
objects (GLOs) at $z=0$ at their virial radius ($\delta \rho/\rho \approx 200$) and 
consider only 
those with more than
250 baryonic particles within their virial radius. 

In hierarchical clustering scenarios, the SF process in GLOs
is affected  by mergers
and interactions (Tissera 2000,
see also observational evidence 
in Barton et al. 1999). 
In our simulations, we found that SF rate histories
can be described as series of starbursts supperposed to a continuous 
component.

For each GLO we can also follow the
evolution of the metallicity of the stellar component 
and calculate the 
age-metallicity  relation (AMR;
e.g. 
Rocha-Pinto et al. 2000).
The mean AMRs estimated for our GLOs 
show the expected trend with high metallicity stars
forming at more recent times in agreement with observations.
The  dispersion found in the simulations indicates 
the existence of coeval SF sites of different metallicities 
at a given time.  
The  values and trend of these relations depend on
the particular evolutionary history of each GLO,
and the SF and SN model parameters
(see Tissera et al. 2000 for details). 

The [O/Fe] versus [Fe/H] is an important observational constrain for chemical
models and may give information on the chemical history of our Galaxy.
We estimated this relation for our GLOs.
\begin{figure}[h]
\caption
{The rate of [O/Fe] as a function of [Fe/H] for
GLO 596 in experiments
S2, S3, S4 and S5. 
Large circles represent observational data
(Gratton et al. 1996).}
\end{figure}
Figure 1 shows a typical example where we can see 
how the SN and SF parameters affect the distributions. 
These abundance relations also depend on the 
the evolutionary history of each GLO.

The chemical abundances in the ISM in gas-rich galaxies
allows to trace the evolution of individual galaxies, 
being HII regions and early B-type main sequence objects
the most accesible probes of interstellar
abundances.   Fairly steep negative
gradients are found: 
$-0.07 \pm 0.01$ dex Kpc$^{-1}$ within $6 < R_{\rm G} < 18$ Kpc
for oxygen 
(Smartt \& Rolleston 1997).
Figure 2
shows abundance gradients of the oxygen for  
GLOs 596 and 325 as examples.
The curves represent the mass-weighted averages of the metal
abundances at each particle position: they 
clearly show  negative abundance gradients.
\begin{figure}
\plotfiddle{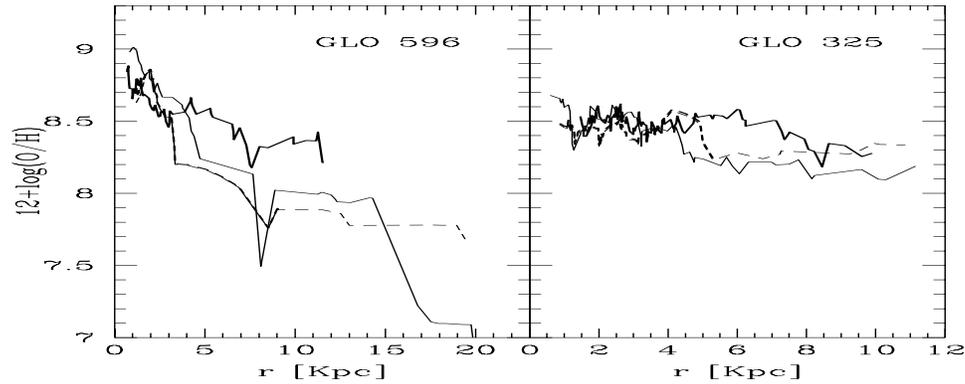}{1.7in}{0}{65}{26}{-190}{-45}
\caption{
Oxygen abundance gradient for gas in hybrid
particles
of GLOs 596 and 325 in experiments 
S1
(thick full line), S2 (thick dashed line),
S3 (thin dashed line) and  S4 (thin full line).}
\end{figure}
The slopes of the calculated gradients range from
$-0.01$ to $-0.4$ dex kpc$^{-1}$,
in agreement with observations.
The differences in the abundance gradient slopes for a given  
GLO are due to the SF mechanism
and SN parameters, and the 
fact that we are only including gas in hybrid particles.

The  slopes  of GLO 325 
are less pronounced 
than those of GLO 596. 
The different behaviour may be due to the fact that
the gas in GLO 596 forms a
well-defined disk, while 
GLO 325 is a clear spheroid, indicating that 
their histories of formation and evolution have been 
very different.

To sum up, in our simulations,
GLOs have different evolutionary history in consistency with a hierarchical
clustering scenario that affect their SF rates and chemical evolution.
This chemical model can take all these physical processes into account, 
resulting in a powerful tool to study galaxy formation.

\end{document}